\documentclass[preprint,prl,showpacs,amsmath,amssymb,preprintnumbers]{revtex4-1}

\usepackage{graphicx}
\usepackage{color}
\usepackage{subfigure}
\usepackage{booktabs}
\usepackage{fancyhdr}

\begin{document}

 \author{Jan Gieseler$^1$, Lukas Novotny$^{2}$ and Romain Quidant$^{1,3}$}

 \affiliation{$^{1}$ ICFO-Institut de Ciencies Fotoniques, Mediterranean Technology Park,
 08860 Castelldefels (Barcelona), Spain}

 \affiliation{$^{2}$ Photonics Laboratory, ETH Z\"urich, 8093 Z\"urich, Switzerland}
 \affiliation{$^{3}$ ICREA-Instituci{\'o} Catalana de Recerca i Estudis Avan\c{c}ats, 08010 Barcelona, Spain}

\date{\today}
\title{Thermal nonlinearities in a nanomechanical oscillator}

\begin{abstract}
{\bf Nano- and micromechanical oscillators with high quality (Q) factors have gained much attention for their potential application as ultrasensitive detectors. In contrast to micro-fabricated devices, optically trapped nanoparticles in vacuum do not suffer from clamping losses, hence leading to much larger $Q$-factors. We find that for a levitated nanoparticle the thermal energy suffices to drive the motion of the nanoparticle into the nonlinear regime.
First, we experimentally measure and fully characterize the frequency fluctuations originating from thermal motion and nonlinearities. Second, we demonstrate that feedback cooling can be used to mitigate these fluctuations.
The high level of control allows us to fully exploit the force sensing capabilities of the nanoresonator.
Our approach offers a force sensitivity of $20\,\rm zN\,{Hz}^{-1/2}$, which is the highest value reported to date at room temperature, sufficient to sense ultra-weak interactions, such as non-Newtonian gravity-like forces.}
\end{abstract}

\maketitle

Recent developments in optomechanics have evolved toward smaller and lighter resonators featuring high quality (Q) factors, which are important for the sensing of tiny masses \cite{Chaste:2012kn,Yang:2006cl}, charges \cite{Cleland:1998uc}, magnetic fields \cite{Rugar:2004wi} and weak forces \cite{Stipe:2001fl,Moser:2013go}.
The presence of a force field or the adhesion of a small mass induces a change in the mechanical response and can be monitored by tracking either the oscillation frequency, phase or its amplitude. Ultimately, dissipation losses as well as  thermomechanical noise and temperature fluctuations limit the $Q$-factors of clamped oscillators and consequently their sensing performance  \cite{Postma:2005ez,Cleland:2002jz,Ekinci:2004eo}. This can be circumvented by using an optically trapped nanoparticle in high vacuum. Indeed, the $Q$-factor of a levitated particle is only limited by collisions with residual air molecules and  can potentially reach  $10^{12}$ for small particles in ultra high-vacuum \cite{Ashkin:1971dd,Gieseler:2012bi,Li:2011jl,RomeroIsart:2010iv,Chang:2010jn}. In this letter we first show that an optically trapped nanoparticle is sufficiently sensitive that thermal forces drive it out of its linear regime. Additionally, we demonstrate that feedback cooling can be used to mitigate frequency fluctuations associated with the thermal nonlinearities thereby recovering the force sensing capabilities of the oscillator.

\begin{figure}[!b]
 \begin{center}
\includegraphics[width=0.6\textwidth]{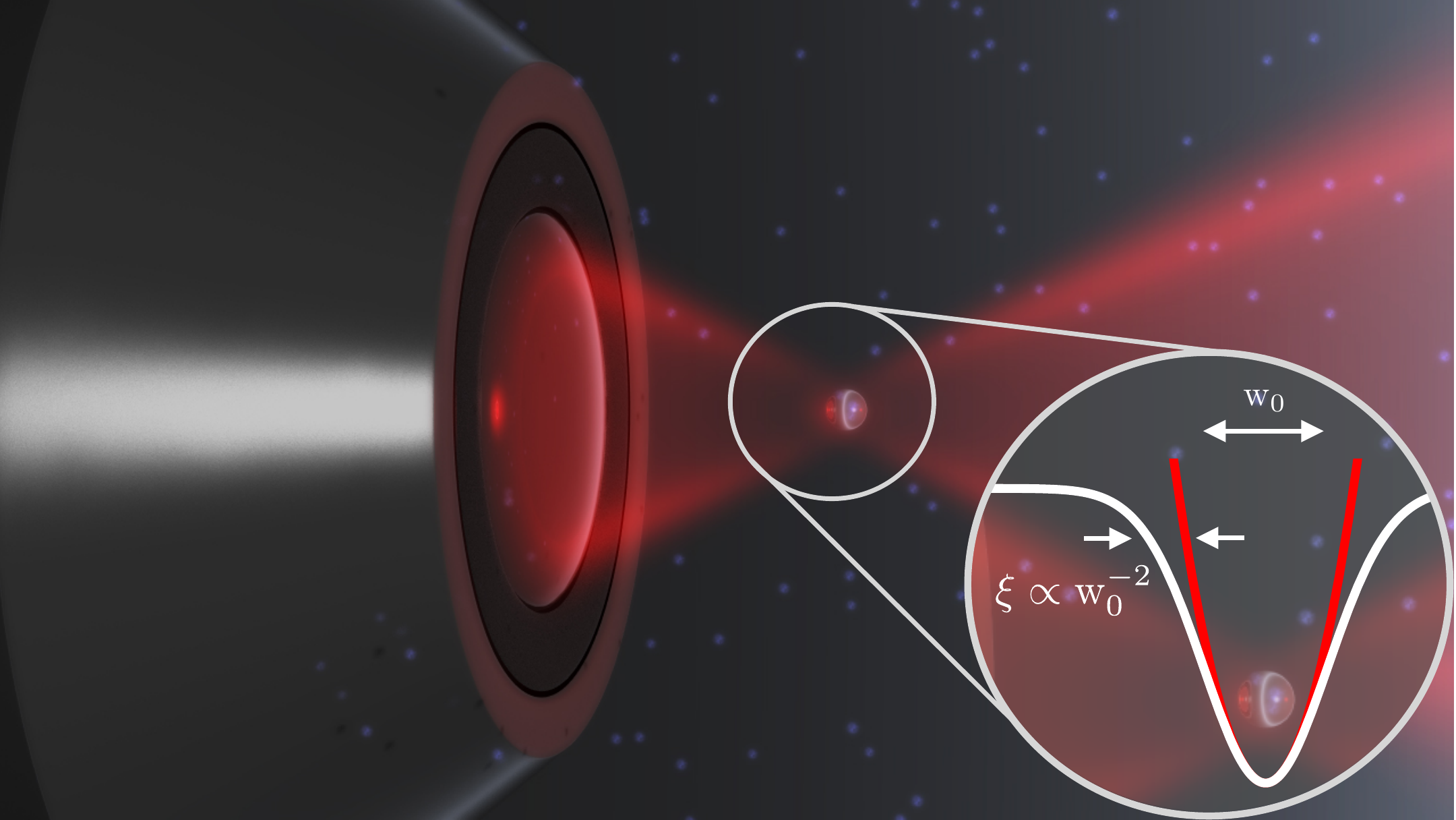}
\end{center}
\caption{{\bf Experimental configuration}
A silica nanoparticle is trapped by a tightly focused laser beam. Random collisions with residual air molecules drive the particle into the nonlinear regime of the potential.
(inset) The focal intensity distribution forms a trap which can be approximated by a Gaussian potential (white). The deviation from a harmonic potential (red) is described by a Duffing nonlinearity.
\label{fig:ExperimentalSetup}}
\end{figure}

In our experiment, a silica nanoparticle with diameter $\sim 75\rm nm$ is trapped in the focal region of a  tightly focused NIR laser beam ($\lambda=1064\rm nm$, polarized along the x-axis). The intensity near the focus of the objective (${\rm NA}=0.8$) can be well approximated by Gaussian functions (c.f. Fig.\ref{fig:ExperimentalSetup}), defining a focal volume of  $w_x\times w_y\times w_z\approx 0.69\rm \mu m\times 0.54\rm \mu m\times 1.36\rm \mu m$ \footnote{estimated from numerical calculation of a highly focused beam with NA=0.8}. For large displacements, the optical potential becomes anharmonic featuring a  Duffing nonlinearity. For a Gaussian field distribution the nonlinear coefficients are given by 

\begin{equation}\label{eqn:NonlinearCoefficients}
\xi_j=-2/{\rm w_j}^2.
\end{equation}

For small displacements $|x_i|\ll |\xi_j^{-1/2}|$, the nonlinearity is negligible and the three motional degrees of freedom decouple. Due to the asymmetry of the optical focus, the oscillation frequencies $\Omega_i=(k_i/m)^{1/2}$ along the three major axes are different ($\Omega_z/2\pi=37\rm kHz,\;\Omega_x/2\pi=125\rm kHz,\;\Omega_y/2\pi=135\rm kHz$). The linear trap stiffness is given by $k_i=\alpha E_0^2/{\rm w_i^2}$, where $E_0$ is the electric field intensity at the focus, ${\rm w_i}$ is the beam waist radius or Rayleigh range. The gradient of the optical intensity distribution exerts a restoring force $F_i^{\rm grad}=-k_i \left(1+\sum_{j=x,y,z} \xi_{j} x_j^2\right)x_i$ on a dipolar particle with polarizability $\alpha$, that is displaced from the trap center by $x_i$. For a sphere of radius $a$ and dielectric constant $\epsilon_p$, the polarizability is  $\alpha=4\pi a^3\epsilon_0(\epsilon_p-1)\left/(\epsilon_p+2)\right.$, $\epsilon_0$ being the vacuum permittivity.\\

We experimentally determine the nonlinear coefficients by parametric excitation through modulation of the trapping laser at a frequency close the parametric resonance $\Omega_{\rm mod}\approx 2\Omega_i$ (supplementary information)
and find $(\xi_z,\xi_x,\xi_y)=(-1.11,-7.43,-8.86)\mu m^{-2}$, in good agreement with the values estimated from the size of the focus \eqref{eqn:NonlinearCoefficients}.\\

The equation of motion for each spatial degree of freedom ($i=x,y,z$) is given by

\begin{equation}\label{eq:EoMDuffing}
\ddot{x_i} + \Omega_i Q^{-1}_i \dot{x_i}+ \Omega_i^2 \left(1+\sum_{j=x,y,z} \xi_{j} x_j^2\right)x_i=\mathcal{F}_{\rm fluct}\left/m\right..
\end{equation}

In the following we concentrate on a single degree of freedom and denote the corresponding resonance frequency by $\Omega_0$ and the quality factor by $Q$. Random collisions with residual air molecules provide both damping $\Gamma_0=\Omega_0Q^{-1}$ and stochastic excitation $\mathcal{F}_{\rm fluct}$ of the trapped nanoparticle. From kinetic theory we find that the damping coefficient of a particle in a rarified gas is given by~\cite{Epstein:1923td,Chang:2010jn}

\begin{equation}\label{eqn:DampingCoefficient}
\Gamma_0=\frac{64 a^2 }{3m\bar{v}}P,
\end{equation}

where $\bar{v}=\left(8k_B T/\pi \mu\right)^ {1/2}$ is the average velocity and $\mu$ is the weight of the air molecules~\cite{OHanlon:49s1Dwu-}.
The random force $\mathcal{F}_{\rm fluct}$ is related to the damping coefficient by the fluctuation-dissipation theorem $\langle \mathcal{F}_{\rm fluct}(t)\mathcal{F}_{\rm fluct}(t')\rangle=2m \Gamma_0 k_B T\delta(t-t')$. The damping coefficient determines the frequency stability of the harmonic oscillator $\Delta \Omega_{\rm L}=\Omega_0Q^{-1}$, and the temperature-dependent stochastic excitations determine the minimum oscillation amplitude according to $r_{\rm th}=\sqrt{k_B T\left/m\Omega_0^2\right.}$.
The thermal amplitude $r_{\rm th}$ is usually small compared to the dimensions of the oscillator. However, for a small and hence light oscillator like our levitated nanoparticle, the thermal amplitude eventually becomes comparable to the particle size. Consequently, a proper description of the particle motion requires the inclusion of nonlinearities. The latter give rise to a frequency shift $\Delta \Omega_{\rm NL}=3\xi\Omega_0/8\, r_{\rm th}^2$ \cite{Anonymous:JsaU-zBe,Lifshitz:2008td}. In contrast to linear thermal frequency fluctuations, {\em nonlinear} frequency fluctuations add frequency noise but do not affect the damping.\\

In order to resolve the nonlinear frequency shift originating from thermal motion,  the nonlinear contribution must be larger than the linear one, that is
\begin{equation}\label{eqn:ThermalNonlinearCondition}
\mathcal{R}=\frac{\Delta\Omega_{\rm NL}}{\Delta\Omega_{\rm L}}=\frac{3\xi Q k_BT}{8 \Omega_0^2 m}\gg 1,
\end{equation}
where $T$ is the temperature of the residual gas and $k_B$ is Boltzmann's constant. To fulfill condition \eqref{eqn:ThermalNonlinearCondition} a light  and high-$Q$ mechanical resonator is required. In our experiment, $m=3\times 10^{-18} \rm kg$ and $Q=10^8$, as determined in a ring-down measurement at  a pressure of $P=0.5\times 10^{-6}\rm mBar$. These parameters place us well into the nonlinear regime. Importantly, the dependence of the $Q$-factor on pressure $P$ allows us to continuously tune the system between the linear and nonlinear regimes.\\

\begin{figure}[!t]
 \begin{center}
\includegraphics[width=0.8\textwidth]{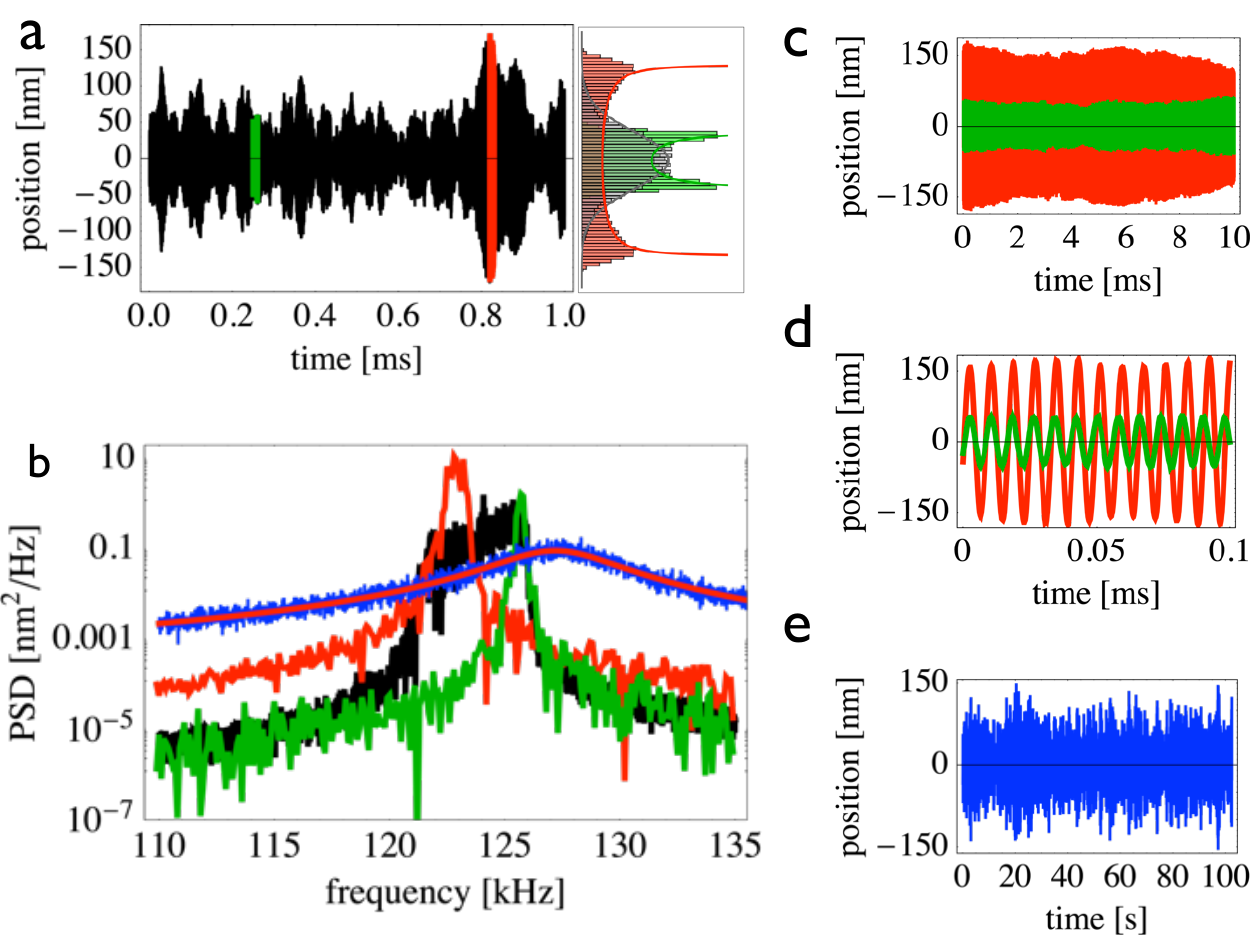}
\end{center}
\caption{{\bf Nonlinearity-induced frequency fluctuations}
(a) Time trace of the particle motion along $x$ at $1.2\times 10^{-2}\,\rm mBar$. The oscillation amplitude changes randomly and the positions are normally distributed. However, on time scales short compared to the relaxation time $\tau$, the particle motion is sinusoidal with a constant oscillation amplitude over many cycles (c,d).
(b) From the long time trace (black) and the short time traces (red, green) we calculate the power spectral density (PSD). For short observation times we observe a Fourier limited symmetric PSD with an amplitude dependent center frequency. The overall PSD (black) results from a temporal average of the instantaneous PSDs (red, green). In contrast, the PSD of a low Q oscillator (blue) is described by a Lorentzian peak at all times (red thin line).
(e) Time trace of particle motion at $6\,\rm mBar$ used to calculate the (blue) Lorentzian PSD in subfigure (b).
\label{fig:TimetracesAndSpectra}}
\end{figure}

To demonstrate the differences between a thermally driven harmonic oscillator ($\mathcal{R}\ll1$) and an anharmonic oscillator ($\mathcal{R}\gg 1$), we compare the particle's motion at high pressure ($6\,\rm mBar$) and at low pressure ($1.2\times 10^{-2}\,\rm mBar$). These pressures correspond to Q-factors of $25$ and $12\times 10^3$, respectively. At high pressures (low Q) the dominant source of frequency fluctuations is linear damping $\Delta\Omega_{\rm L}=\Gamma_0$ and the power spectral density (PSD) of the particle motion features a single symmetric Lorenzian peak, whose width is equal to the linear damping coefficient $\Gamma_0$ (Fig. \ref{fig:TimetracesAndSpectra}b). In contrast, at low pressure (high Q) nonlinear frequency fluctuations $\Delta\Omega_{\rm NL}=3\xi\Omega_0\left/8\right.\,r_{\rm th}^2$ dominate and we observe an asymmetric peak that is considerably broader than what is expected for the equivalent linear oscillator. However, if we limit the observation time to time intervals $1/\Delta\Omega_{\rm NL}<\tau<1/\Delta\Omega_{\rm L}$, we find a clean oscillation with an almost constant amplitude, corresponding to a narrow and symmetric peak in the frequency domain. For large oscillation amplitudes the peak appears down-shifted, consistent with the measured negative Duffing nonlinearity.
Consequently, for observation times $\gg 1/\Delta \Omega_{\rm NL}$, the non-Lorenzian peak becomes a weighted average \cite{Dykman:1990te}

\begin{equation}
S_{\rm NL}(\Omega)=\int \rho(E) S_{\rm L}(\Omega,E) dE,
\end{equation}
over Lorenzian peaks centred at the shifted frequency $\hat{\Omega}_0(E)=\Omega_0+3\xi/(4m \Omega_0)\, E$ and weighted by the Gibbs distribution $\rho(E)=Z^{-1}\exp(-E/k_BT)$.
Here, $Z=\int \rho(E)dE$ is the partition function and 

\begin{equation}
S_{\rm L}(\Omega,E)=\frac{E}{\pi m \Omega_0^2}\frac{\Gamma_0}{\left(\Omega-\hat{\Omega}_0(E)\right)^2+(\Gamma_0/2)^2}.
\end{equation}
is the power spectral density of a harmonic oscillator with frequency $\hat{\Omega}_0(E)$ and energy $E$.\\

\begin{figure}[!t]
 \begin{center}
\includegraphics[width=0.8\textwidth]{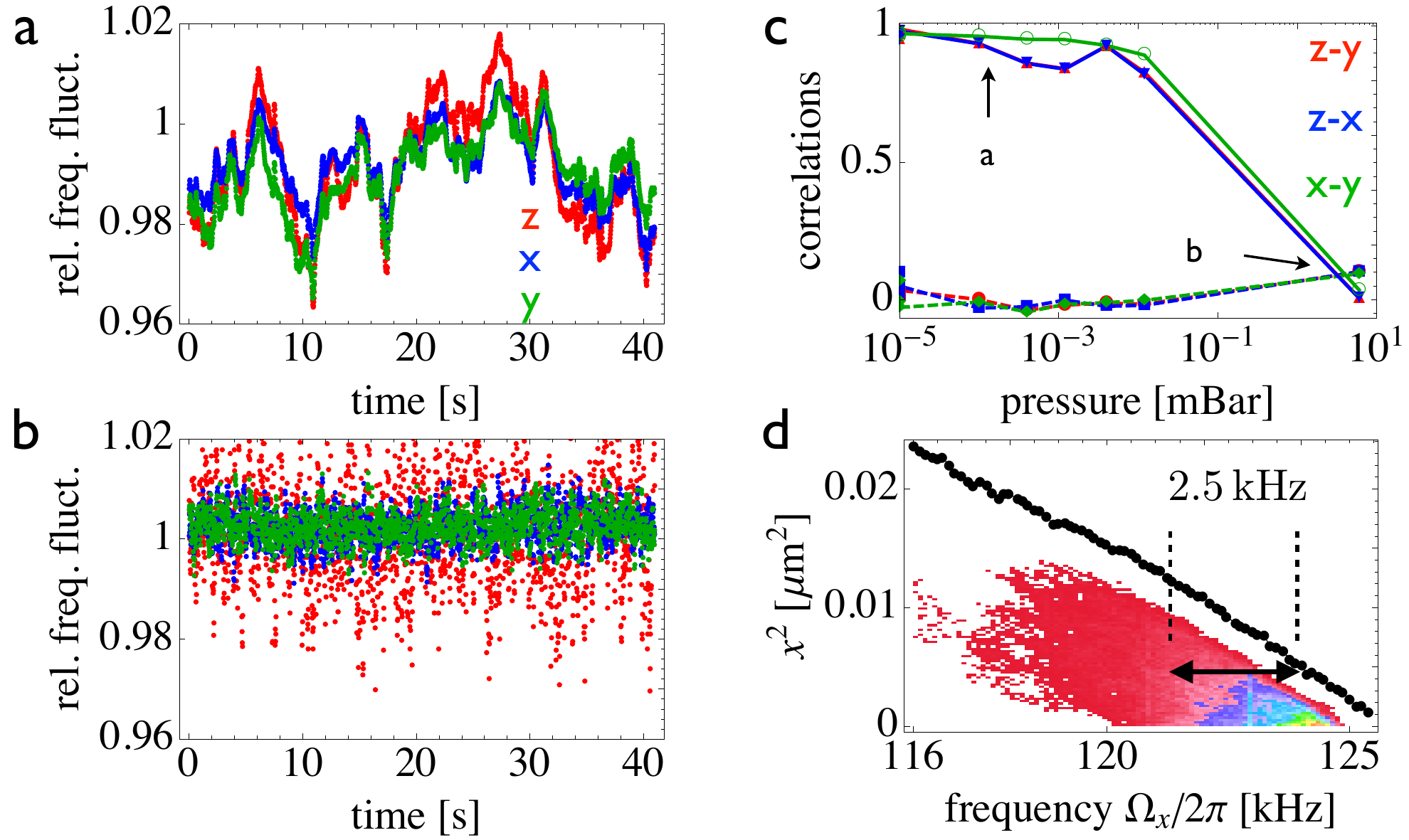}
\end{center}
\caption{{\bf Frequency and energy correlation.}
(a) Time traces of the instantaneous oscillation frequencies $\Omega_x$, $\Omega_y$ and $\Omega_z$ evaluated at low pressure and (b) at high pressure.
(c) Frequency and energy correlations as a function of pressure.
The energies (dashed, lower curves) are uncorrelated and independent of pressure.
The frequencies (solid, upper curves) are correlated at low pressure where nonlinear fluctuations dominate and uncorrelated at high pressure, where linear fluctuations dominate.
The points labeled 'a' and 'b' indicate the pressures at which figures (a) and (b) were evaluated.
(d) Energy vs. frequency for the oscillator in x-direction ($x$ mode) obtained from both random fluctuations (colored density plot) and from parametric excitation (black).
The thermal excitation of the orthogonal modes ($y,z$) shifts the resonance by $\approx 2.5\,\rm kHz$ in good agreement with the value estimated from the thermal amplitude and the measured nonlinear coefficients.
\label{fig:Correlations}}
\end{figure}

To quantify the frequency fluctuations, we continuously measure the instantaneous energy $E_i(t_j)$ and frequency $\Omega_i(t_j)$ of the three spatial modes ($i=x,y,z$), which are calculated from position time traces $x^{(j)}_i(t)$ (where $t_j-\tau/2<t< t_j+\tau/2$) of $\tau=20\, \rm ms$ duration. Analysing the correlations between the instantaneous frequencies and energies, we verify that the frequency fluctuations are due to nonlinearities in the optical potential. Figure~\ref{fig:Correlations} shows the correlations between $E_i$ and  $\Omega_i$ as a function of pressure, calculated from $30\rm min$ long time traces. The nonlinearity is conservative and, thus, doesn't change the particle energy, which is determined only by random molecule collisions. Therefore, the energy of the three degrees of freedom are uncorrelated. In contrast, a change in energy of one mode shifts the frequency of all modes (Eq. \eqref{eq:EoMDuffing}). At low pressure ($\mathcal{R}>1$), the nonlinearities dominate and the frequency fluctuations are highly correlated. In contrast, at high pressure ($\mathcal{R}<1$), linear damping dominates and consequently the frequencies become uncorrelated.

In Figure \ref{fig:Correlations}d we plot the oscillator energy as a function of the oscillator frequency at a pressure of $10^{-5}$ mBar. We plot both the thermally-induced dependence (density plot) and the dependence resulting from external parametric modulation of the trap potential (dots, Supplementary Information). As expected, in both cases the energy scales linearly with the frequency. However, because of nonlinear frequency fluctuations, the data of the thermally driven oscillator cover a broad range and appear downshifted with respect to the data of the parametrically-driven oscillator. Indeed, according to eq.~\eqref{eq:EoMDuffing}, thermal excitation of the orthogonal modes ($y$ and $z$) also shifts the frequency of the mode under consideration (here $x$). Therefore, for a fixed amplitude of $x$, we observe a distribution of frequencies. In contrast, the response of the driven oscillator is sharp because feedback cooling keeps the orthogonal modes at a low amplitude,  while we parametrically excite only the mode under consideration. The shift of the center of the frequency distribution ($\approx 2.5\rm kHz$) is in good agreement with the value estimated from the measured nonlinear coefficients and average thermal amplitudes. 

\begin{figure}[!t]
 \begin{center}
\includegraphics[width=0.8\textwidth]{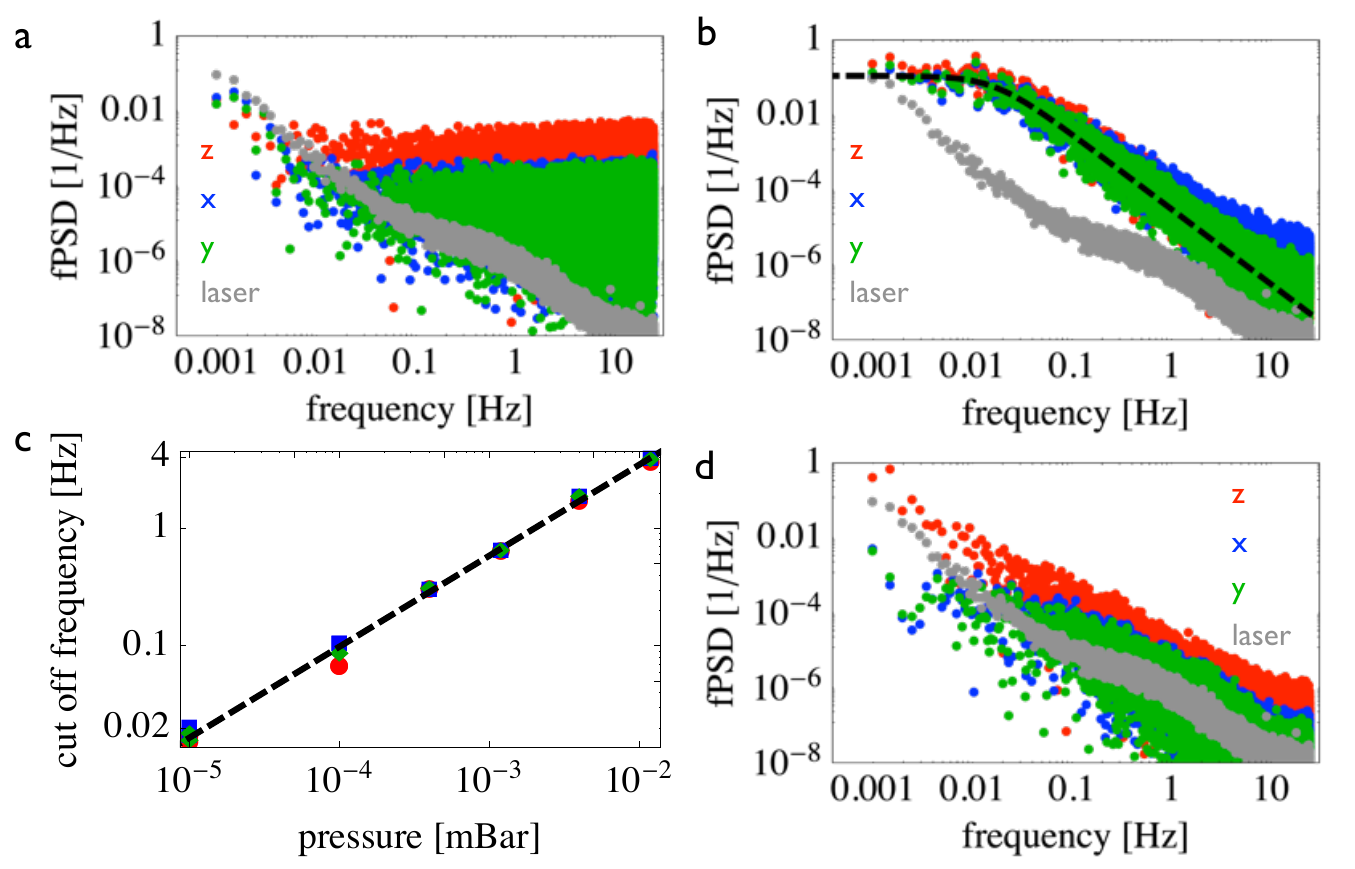}
\caption{{\bf Pressure dependence of frequency fluctuations.}
Pressure dependence of frequency fluctuations.
(a) At high pressures (6mBar), the fluctuations are solely determined by the linear damping (low Q).
(b) When the Q factor becomes larger (1e-5mBar), the fluctuations are dominated by nonlinear amplitude-frequency conversion.
In this case, the frequency power spectral density (fPSD) exhibits a characteristic cut-off, which we extract from a fit to \eqref{eqn:NonlinearfPSD} (black dashed line).
(c) The characteristic cut-off frequency depends on the Q factor which scales linearly with pressure.
(d) Using feedback cooling, the fluctuations can be suppressed to the level of the laser intensity fluctuations (gray).
\label{fig:PressureDependence}}
\end{center}
\end{figure}

In figure \ref{fig:PressureDependence} we show the power spectral density of the relative frequency $\Omega/\Omega_0$ (fPSD) as a function of pressure.
The fPSDs are calculated from $30\,\rm min$ long timetraces of the instantaneous frequencies (c.f. Fig.\ref{fig:Correlations}).
For low $Q$, the fPSD is flat as expected for a harmonic oscillator.
In contrast for high $Q$  the nonlinear coupling maps the Lorenzian power spectral density of the amplitude onto the frequency power spectral density, which is therefore given by 
\begin{equation}\label{eqn:NonlinearfPSD}
S_{\rm f}(\Omega)=I\frac{\Omega_c/\pi}{\Omega^2+\Omega_c^2},
\end{equation}
where $I$ is the total spectral power, which is independent of pressure. 
The characteristic cut-off frequency $\Omega_c$ has a clear pressure dependence, as shown in Fig. \ref{fig:PressureDependence}c.
This further confirms that the fluctuations arise as a combination of nonlinearities and thermal excitations.

The small mass and high Q-factor make the levitated nanoparticle an ultrasensitive force sensor with sensitivity of $S_F=4k_BT m\Omega_0/Q\approx (20\rm zN)^2/{\rm Hz}$ at room temperature.
This is the highest value reported to date and compares to the best values achieved at cryogenic temperatures \cite{Moser:2013go}.
In practice though, nonlinear effects lead to frequency fluctuations in ultra-high Q oscillators, which are typically considered detrimental to the oscillator performance.
It is possible to surpass this limit by operating the oscillator at special points \cite{Villanueva:2013cv}.
In the following we show that the nonlinear frequency fluctuations can also be suppressed by feedback cooling \cite{Gieseler:2012bi}.
Feedback cooling lowers the oscillation amplitude and therefore reduces the thermal motion of the oscillator.
Under the action of feedback, the effective temperature reduces to $T_{\rm eff}=(\Gamma_0/\Gamma_{\rm fb})T$, where $\Gamma_{\rm fb}$ is the total damping with feedback \cite{Mertz:1993cd,Gieseler:2012bi}.
As shown in Fig.\ref{fig:PressureDependence}d, we manage to reduce the frequency fluctuations to the level of the laser intensity fluctuations, which become the main source of frequency noise. Using active stabilization techniques, laser noise can by brought to the level of  $10^{-8}/\sqrt{\rm Hz}$ \cite{Seifert:2006wp}.\\

\begin{figure}[!t]
\begin{center}
\includegraphics[width=0.8\textwidth]{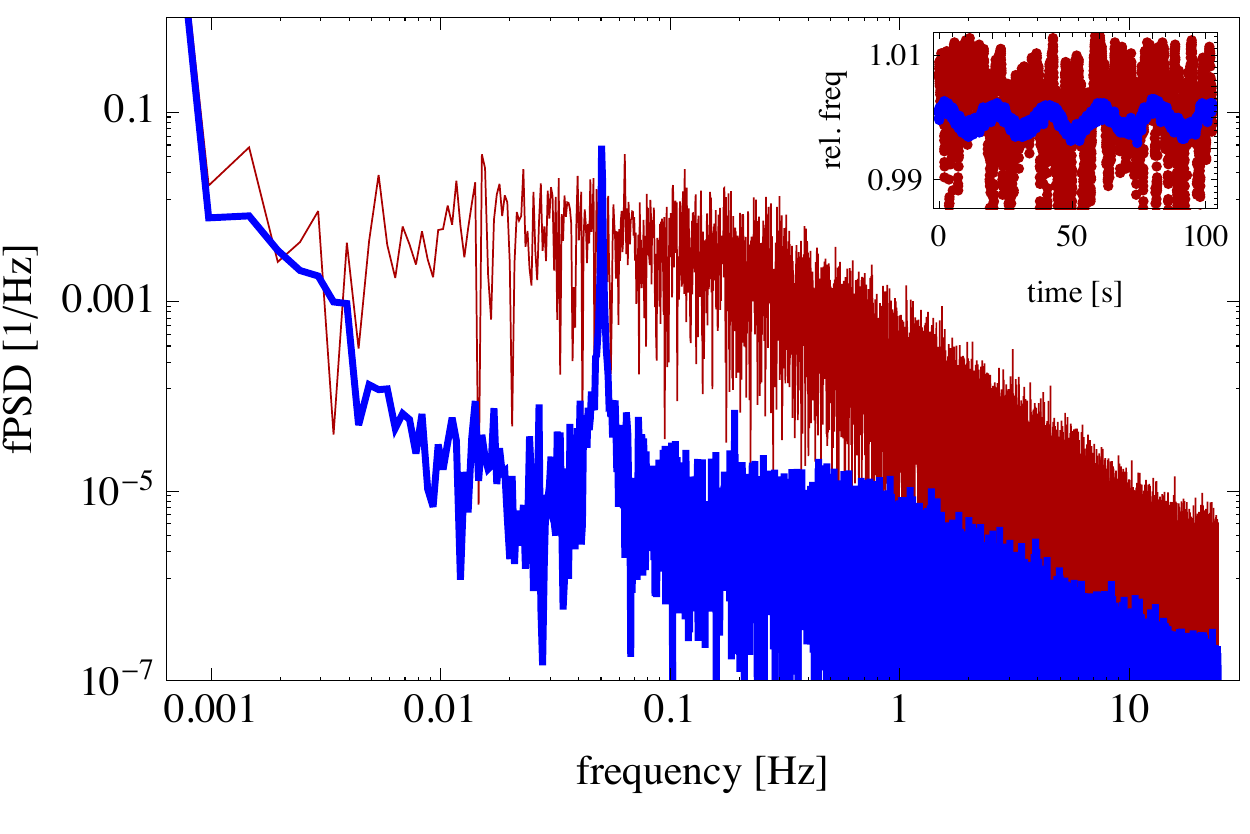}
\caption{{\bf Detection of a Periodic force gradient using feedback cooling.}
A periodic potential landscape is emulated by modulating the trapping frequency at $50\,\rm mHz$.
In absence of feedback cooling the small signal is overwhelmed by noise (red).
Feedback cooling reduces the random frequency fluctuations thereby making it possible to detect the signal (blue).  (inset) Time domain signal of relative frequency with (blue) and without feedback (red).
\label{fig:ModulationGradient}}
\end{center}
\end{figure}

We demonstrate the improved sensitivity by mimicking a periodic potential landscape. A modulation at $50\rm mHz$ is applied to the trapping laser. The modulation causes a variation of the force gradient, which is measured as a frequency shift. As shown in figure~\ref{fig:ModulationGradient}, without feedback the signal is overwhelmed by noise while with feedback, the fluctuations are suppressed down to the level of laser intensity fluctuations and the applied signal is clearly visible. Using feedback cooling, we are able to improve the sensitivity of the oscillator by two orders of magnitude and achieve differential frequency resolutions of $\partial\Omega_0/\partial\Omega$ of $3\times 10^{-3}/\sqrt{\rm Hz}$ for frequencies below 1 Hz and $1\times 10^{-4}/\sqrt{\rm Hz}$ for frequencies larger than $10\rm Hz$. In the absence of laser intensity noise the highest sensitivity is obtained when linear and nonlinear fluctuations contribute equally. Since feedback cooling reduces both the effective temperature $T_{\rm eff}$ and the effective quality factor $Q_{\rm eff}=\Omega_0/\Gamma_{\rm fb}$, the optimum feedback gain is $Q_{\rm eff}^{\rm (opt)}=\left(8 m \Omega_0^2 Q\left/3\xi k_BT\right.\right)^{1/2}$, for which the minimum frequency shift is given by (Supplementary information)

\begin{equation}\label{eqn:MinFrequencyShift3}
|\partial\Omega_0/\Omega_0|_{\rm min}=\sqrt{\frac{ B}{ \Omega_0 Q_{\rm eff}^{\rm (opt)} }},
\end{equation}

where $B$ denotes the bandwidth. For the values presented here we obtain $3\times 10^{-6}$, sufficient to sense ultraweak interactions, such as non-Newtonian gravity-like forces \cite{Geraci:2010it}.

The ultimate cooling limit is defined by the zero point motion  $r_{\rm zp}=\sqrt{\hbar\left/2m\Omega_0\right.}$. In order to resolve the nonlinear frequency shift due to the zero point motion $\Delta\Omega_{\rm zp}$, the condition

\begin{equation}
\frac{\Delta\Omega_{\rm zp}}{\Delta\Omega_{\rm L}}=\frac{3}{8}Q \xi r_{\rm zp}^2\gg 1
\end{equation}

has to be satisfied in analogy to \eqref{eqn:ThermalNonlinearCondition}. This requires a $Q$ factor of $Q=\Omega_0/\Delta\Omega_{\rm zp}\approx  10^{10}$. In absence of other noise sources, this regime is reached for pressures below $10^{-8} \rm mBar$. 

In conclusion, we have demonstrated that a laser-trapped  nanoscale particle in high vacuum defines an ultrasensitive force sensor. The thermal motion of the residual gas drives the nanoparticle into its nonlinear regime, which gives rise to frequency fluctuations. Using a parametric feedback cooling scheme, we can stabilize the nanoparticle and suppress its nonlinearities, without sacrificing sensitivity. We expect that feedback-controlled nanoparticles will find applications for sensing a wide  range of interactions, including van der Waals and Casimir forces \cite{Geraci:2010it}, nuclear spins \cite{Rugar:2004wi}, and gravitation \cite{Arvanitaki:2013ei}. \\

This research was funded by ETH Zurich, Fundaci\'o Privada CELLEX, and ERC-Plasmolight (No. 259196). We thank Adrian Bachtold and Marko Spasenovic for valuable input and help.


\end{document}